\documentstyle[12pt,epsfig]{article}

\input epsf

\newcommand{\frat}[2]{\frac{\textstyle #1}{\textstyle #2}}

\newcommand{\vf}[1]{\mbox{\boldmath $#1$}}

\begin{document}
\begin{center}
{\Large {\bf
Role of quark-instanton liquid interactions\\
 in colour  superconductivity phase}}\\
\vspace{0.5cm}
S. V. Molodtsov, G. M. Zinovjev$^\dagger$\\
\vspace{0.5cm}
{\small\it State Research Center,
Institute of Theoretical and Experimental Physics,
117259, Moscow, RUSSIA}\\
$^\dagger$
{\small \it
Bogolyubov Institute for Theoretical Physics,\\
National Academy of Sciences of Ukraine,
UA-03143, Kiev, UKRAINE}
\end{center}
\vspace{0.5cm}
\begin{abstract}
The interaction of light quarks and instanton liquid is analyzed at finite
density of quark/baryon matter and in the phase of nonzero values of diquark
(colour) condensate. It is shown that instanton liquid perturbation produced
by such an interaction results in an essential increase of the critical value
of quark chemical potential $\mu_c$ which provokes the perceptible increase of
quark matter density around the expected onset of the colour
superconductivity phase.
\end{abstract}
\vspace{0.5cm}

\\
\\

\vspace{0.5cm}
\noindent
The intriguing history of studying the structure of strongly interacting matter
abounded in many interesting but somewhat contradictory phenomenological
results up to the advent of quantum chromodynamics (QCD). Since the time of its
discovery the possible existence of quark-gluon plasma (QGP) phase is one of
the most fundamental predictions of QCD. This phase is phenomenologically
understood as a matter state where quarks and gluons being under extreme
conditions of high temperature (T) and(or) quark/baryon density (driven by the
corresponding chemical potential $\mu$) are able to propagate relatively freely
over the considerable (of course, on the hadron scale) distances. Even the
simplest quantitative estimates of QGP physical characteristics were so
verisimilar that they allowed to launch the process of QGP experimental study
in ultrarelativistic heavy ion collisions. And although the first series of
experiments at SPS CERN and AGS BNL have provided us with encouraging
results a new generation of experiments is necessary to make convincing
conclusions. The physical programme for new experimental research certainly
needs serious quantitative analysis of the QCD phase diagram.

It seemed lattice QCD could be able to resolve the problem starting even on
the first theoretical principles. However, up to now this approach was
effective and apparently reliable in producing the results mainly around the
temperature axis of the ($\mu$---T)-plane and, unfortunately, relies on the
field theory formulation in imaginary time. Nevertheless, lattice QCD has
corroborated the phenomenological expectations that its phase structure is
quite rich \cite{karsh}. In particular, nowadays we know that the chiral
symmetry of initial QCD Lagrangian becomes broken and the quarks are getting
to be confined when the temperature decreases below a critical value. Moreover,
the lattice calculations show us that both physical phenomena result from the
corresponding phase transitions and their critical temperatures practically
coincide. The latter observation indicates that both fundamental phenomena
could be tightly knitted and it should heuristically play an important role to
illuminate the confinement mechanism.

The features of the QCD matter at high baryon densities remains poorly
understood in spite of the recent splash of renewed activity \cite{0}. One of
the major theoretical reasons comes from serious technical difficulties of
lattice simulations with Monte-Carlo techniques due to the complex fermion
determinant in QCD at finite quark densities. From the phenomenological side
the recent experimental data are not very informative for developing the
theory  along this direction. It is clear that present and future experiments with
heavy ions at higher and higher energies should produce the strong interacting
matter rather at higher temperature and closer to the temperature axis of the
($\mu$---T)-plane. In such a situation the phenomenological treatment of
studying along the $\mu$-axis should basically rely on the astrophysical
observations of the compact stars. In the meantime, it was demonstrated that
the diquark pairing in colour anti-triplet attractive channel induced by the
instanton interaction vertex should be much more effective than an one-gluon
exchange and at sufficiently large quark/baryon densities and small T could
lead to colour superconductivity and a sophisticated picture of the QCD phase
diagram \cite{0}.

The present paper continues our preceding one \cite{mz} where the light quark
interactions with the instanton liquid (IL) at nonzero values of quark/baryon
chemical potentials were investigated in the phase of broken chiral symmetry
within the improved method of calculating the generating function of the IL
theory \cite{we0}. The new treatment of the functional integral is mainly
related to an idea t take into account the IL back-reaction upon the quark presence
which was always considered negligible. In fact, this effect is pretty weak
and should manifest itself in the leading order of expansion in the effective
coupling. Nevertheless, it was qualitatively argued that the interaction of IL
and quarks could increase the quark matter density around the onset of
expected colour superconductivity. As known \cite{diakcar}, \cite{rapp} this
density occurred to be surprisingly small ($n_q\sim 0.062~{\mbox{fm}}^{-3}$)
and, hence, the diquark phase would come to play already at the density of
nuclear matter ($n_n\simeq 0.45~{\mbox{fm}}^{-3}$ and quark matter density is
taken to be three times larger than the conventional nuclear one).

Here we obtain the quantitative estimate of the corresponding critical
chemical potential $\mu_c$ analyzing the system of the Gorkov equations for
the colour superconductor which we further calculate. We do not introduce any
simplification of the initial Lagrangian trying to deal with the 'exact'
four-quark interaction which is generated by (anti-)instantons. Such an
intention is simply compulsory because it is dictated by the precision of
problem where the effects of the IL perturbation are expected to be of the
same order of magnitude compared to the terms which could usually be
ignored.

Let us point out that in the IL approach \cite{2} the generating
functional has the following factorized form $$ {\cal Z}~=~{\cal
Z}_g~\cdot~{\cal Z}_\psi~. $$ Here the first factor provides us
with information on the gluon condensate whereas the fermion
factor ${\cal Z}_\psi$ serves to describe the quark practice in
the instanton environment \cite{diakcar}, \cite{2}. In what
follows we use the notations of Ref. \cite{mz} where the
dimensionless variables (motivated by the form of interquark
interaction kernel) were introduced, for example, for the chemical
potential it looks like $\mu\to \mu\bar\rho/2$ and for the momenta
$k_i\bar\rho/2 \to k_i,~i=1, \dots, 4$ where $\bar\rho$ is the
average size of pseudoparticle (PP). The quark determinant ${\cal
Z}_\psi$ may be transformed with the auxiliary integration over
the parameter $\lambda$ to the following form (for the colour
$SU(3)$ group with the quarks of two flavours $N_f=2$):
\begin{eqnarray}
\label{1}
&&{\cal Z}_\psi\simeq\int d\lambda~\int D\psi^\dagger D\psi~\exp\left\{
n~V_4\left(\ln\frat{n\bar\rho^4}{\lambda~N}-1\right)\right\}
\times\nonumber\\
&&\times\exp\left\{\int \frat{dk}{\pi^4}~\sum_{f=1}^{2}\psi^\dagger_{f}(k)
~2~(-\hat k-i\hat\mu)\psi_{f}(k)+V\right\}~,
\\
&&V=2\lambda~(\psi^{\dagger L}_1~L_1~\psi_1^{L})
(\psi^{\dagger L}_2~L_2~\psi_2^{L})+
2\lambda~(\psi^{\dagger R}_1~R_1~\psi_1^{R})
(\psi^{\dagger R}_2~R_2~\psi_2^{R})~,\nonumber
\end{eqnarray}
where $\psi_f^{T}=(\psi_f^{R},\psi_f^{L}),~f=1,2$ are the quark
fields with the spinors of definite chirality,
$\psi_f^{L,R}=P_\pm~\psi_f,~P_\pm=\frat{1\pm\gamma_5}{2}$, $n$ is
the IL density, $V_4$ is 4-volume which the system occupies,
$\mu_\nu=({\vf 0},\mu)$ and $N$ is the normalizing factor. For
clarity we take it to be equal a unity but, in principle, it could
play a role of free model parameter. This factor is inessential
for the models with the fixed value of the packing fraction
parameter $n\bar\rho^4$ but in the model where it is admissible
for variation the weak logarithmic dependence on N appears. The
factors $2$ in Eq. (\ref{1}) come with making use of the
dimensionless variables. The term of four-fermion quark
interaction $V$ may be directly expressed with the chiral
components as $$(\psi^{\dagger L}_f~L_f~\psi_f^{L})= \int
\frat{dp_f dq_f}{\pi^8}~\psi^{\dagger L}_{f\alpha_f i_f}(p_f)~
L_{\alpha_f i_f}^{\beta_f j_f}(p_f,q_f;\mu)~\psi_f^{L\beta_f
j_f}(q_f)~,$$ it being known that for the right hand fields the
substitution $L \to R$ should be performed. The kernels
$L_{\alpha_f i_f}^{\beta_f j_f}$ are defined by the functions
$h_i,~i=1, \dots, 4$ and the zero modes (the solutions of the
Dirac equation with the chemical potential $\mu$ in the PP field)
\begin{eqnarray}
h_4(k_4,k;\mu)&=&\frat{\pi}{4 k}
\{(k-\mu-ik_4)[(2k_4+i\mu)f^{-}_{1}+i(k-\mu-ik_4)f^{-}_2]+\nonumber\\
&+&(k+\mu+ik_4)[(2k_4+i\mu)f^{+}_{1}-i(k+\mu+ik_4)f^{+}_2]\}~,\nonumber
\end{eqnarray}
\begin{eqnarray}
h_i(k_4,k;\mu)&=&\frat{\pi~k_i}{4 k^2}
\left\{(2k-\mu)(k-\mu-ik_4)f^{-}_{1}+(2k+\mu)(k+\mu+ik_4)f^{+}_1+
\right.\nonumber\\
&+&\left[2(k-\mu)(k-\mu-ik_4)-\frat{1}{k}(\mu+ik_4)[k_4^{2}+(p-\mu)^2]
\right]f^{-}_{2}+\nonumber\\
&+&\left.
\left[2(k+\mu)(k+\mu+ik_4)+\frat{1}{k}(\mu+ik_4)[k_4^{2}+(p+\mu)^2]
\right]f^{+}_{2}\right\}~,\nonumber
\end{eqnarray}
where $k=|{\vf k}|$ if the spatial components of 4-vector $k_\nu$ are
considered and
$$f_1^{\pm}=
\frat{I_1(z^\pm)K_0(z^\pm)-I_0(z^\pm)K_1(z^\pm)}
{z^\pm}~,\\
f_2^{\pm}=\frat{I_1(z^\pm)K_1(z^\pm)}{z^2_{\pm}}~,~~
z^\pm=\frat{\rho}{2}\sqrt{k_4^{2}+(k\pm\mu)^2}~,
$$
with the modified Bessel functions $I_i,~K_i~(i=0,1)$.
Let us introduce the scalar function $h(k_4,k;\mu)$ related to
three-dimensional component
$h_i(k_4,k;\mu)=h(k_4,k;\mu)~\frat{k_i}{k}~,i=1,2,3$
(we omit the arguments of functions $h_i$ when it does not mislead).
$$L_{\alpha i}^{\beta j}(p,q;\mu)=S^{i k}(p;\mu)
\epsilon^{k l}~U^{\alpha}_{l}~U^{\dagger\sigma}_{\beta}
\epsilon^{\sigma n}S^{+}_{n j}(q;-\mu)~,
$$
$S(p;\mu)=(p+i\mu)^{-}~h^{+}(p;\mu)~,~~
S^+(p;-\mu)=\stackrel{*}h\!\!{^{-}}(p;-\mu)(p+i\mu)^{+}$ where it is valid
for the conjugated function $\stackrel{*}h_\mu(p;-\mu)=h_\mu(p;\mu)$ and
$\epsilon$ is an antisymmetric matrix with $\epsilon_{12}=-\epsilon_{21}=1$.
Here $p^\pm$ and other similar designations are used for the four-vectors
spanned by $\sigma^\pm_{\nu}$-matrices,
$\sigma^{\pm}_\mu=(\pm i{\vf \sigma},1),$ (${\vf \sigma}$ is the three-vector
of the Pauli matrices), for example, $p^\pm=p^\nu\sigma^\pm_{\nu}$ and
$U$ is a matrix of rotations in the colour space.
The similar relations are valid for the right hand components
$$(\psi^{\dagger R}_f~R_f~\psi_f^{R})=
\int \frat{dp_f dq_f}{\pi^8}~\psi^{\dagger R}_{f\alpha_f i_f}(p_f)~
R_{\alpha_f i_f}^{\beta_f j_f}(p_f,q_f;\mu)~\psi_f^{R\beta_f j_f}(q_f)~,$$
with the kernel
$$R_{\alpha i}^{\beta j}(p,q;\mu)=T^{ik}(p;\mu)
\epsilon^{k l}~U^{\alpha}_{l}~U^{\dagger\sigma}_{\beta}
\epsilon^{\sigma n}T^{+}_{n j}(q;-\mu)~,
$$
where $T(p;\mu)=(p+i\mu)^{+}~h^{-}(p;\mu)~,~~
T^+(p;-\mu)=\stackrel{*}h\!\!{^{+}}(p;-\mu)(p+i\mu)^{-}$.
The components of matrices $(p+i\mu)^{\pm}$ and $h^{\mp}(p;\mu)$ commute
because the vector-function ${\vf h}(p)$ is spanned by the vector ${\vf p}$
only. Then the following identities are easily understood
$$T(p;\mu)=S^{+}(p;-\mu)~,~~T^+(p;-\mu)=S(p;\mu)~.$$
In what follows we omit the $\mu$-dependence of matrices $T,~T^+$ as the
chemical
potential enters the matrix $T$ being always positive and the matrix $T^+$
being negative only. Besides, two other identities would also be helpful
$$\sigma_2 T^{T}(p)\sigma_2=T^+(p)~,~~
\sigma_2 T^{+ T}(p)\sigma_2=T(p)~,
$$
where $T^T$ means a transposed matrix.

Dealing with the following averages interesting to study the diquark
condensates (\ref{1})
$$
\langle\psi^{L,R}_{1\alpha i}(p)\psi^{L,R}_{2\beta j}(q)\rangle=
\epsilon_{12}~\epsilon_{\alpha\beta}~\pi^4~\delta(p+q)~F^{L,R}_{ij}(p)~,
$$
$$
\langle\psi^{L}_{f\alpha i}(p)\psi^{\dagger R}_{g\beta j}(q)\rangle=
\delta_{fg}~\delta_{\alpha\beta}~\pi^4~\delta(p-q)~G^{LR}_{ij}(p)~,
$$
and making use of the effective action of Eq. (\ref{1}) one may obtain the
Gorkov equations similar to ones treated in Ref. \cite{rapp}
\begin{eqnarray}
\label{2}
&&\left[G_{0}^{+}(p)\right]^{-1}~F^L(p)-\Sigma^R(p)~G^{LR~T}(-p)=0~,
\nonumber\\
&&\left[G_{0}^{+}(p)\right]^{-1}~G^{LR}(p)-
\Sigma^R(p)~F^{+R~T}(p)=1~,
\nonumber\\
[-.2cm]
\\[-.25cm]
&&\left[G_{0}^{-}(p)\right]^{-1}~F^R(p)-
\Sigma^L(p)~G^{RL~T}(-p)=0~,\nonumber\\
&&\left[G_{0}^{-}(p)\right]^{-1}~G^{RL}(p)-
\Sigma^L(p)~F^{+L~T}(p)=1~,\nonumber
\end{eqnarray}
where $\left[G_0^{\pm}(p)\right]^{-1}=-2~(p+i\mu)^\pm$ means the free Green
function, $\Sigma^{R}(p)=\Delta^{R}~T(p)~\epsilon~T^{T}(-p)$,
$\Sigma^{L}(p)=\Delta^{L}~T^{+}(p)~\epsilon~T^{+T}(-p)$ and
$\Delta^{L,R}$ denotes a gap. The form of $\Sigma$-matrices results from the
kernel structure of equations if averaging over colour orientations done
(remember that we consider colour stochastic ensemble). In order to complete
Eqs. (\ref{2}) we need the following gap equations
$$\epsilon \Delta^{R}=\frat{2\lambda}{N_c(N_c-1)}~
\int \frat{dq}{\pi^4} \left[T^{+}(q)~F^{R}(q)~T^{+T}(-q)-
T^{+}(-q)~F^{R~T}(q)~T^{+T}(q) \right]~,
$$
$$\epsilon \Delta^{L}=\frat{2\lambda}{N_c(N_c-1)}~
\int \frat{dq}{\pi^4} \left[T(q)~F^{L}(q)~T^{T}(-q)-
T(-q)~F^{L~T}(q)~T^{T}(q) \right]~.
$$
Apparently, the right hand sides of these equations are proportional $\epsilon$
because they are the differences a matrix and its transposed form. Similar
equations are valid for the conjugated matrices
\begin{eqnarray}
\label{3}
F^{+L~T}(p)&\left[G_{0}^{-}(p)\right]^{-1}&-G^{RL~T}(-p)~\Sigma^{+R}(p)=0~,
\nonumber\\
G^{RL}(p)&\left[G_{0}^{-}(p)\right]^{-1}&-
F^R(p)~\Sigma^{+R}(p)=1~,
\nonumber\\
[-.2cm]
\\[-.25cm]
F^{+R~T}(p)&\left[G_{0}^{+}(p)\right]^{-1}&-
G^{LR~T}(-p)~\Sigma^{+L}(p)=0~,\nonumber\\
G^{LR}(p)&\left[G_{0}^{+}(p)\right]^{-1}&-
F^L(p)~\Sigma^{+L}(p)=1~,\nonumber
\end{eqnarray}
with the corresponding gap equations
$$\epsilon \Delta^{+R}=\frat{2\lambda}{N_c(N_c-1)}~
\int \frat{dq}{\pi^4} \left[T^{T}(-q)~F^{+R~T}(q)~T(q)-
T^{T}(q)~F^{+R}(q)~T(-q) \right]~,
$$
$$\epsilon \Delta^{+L}=\frat{2\lambda}{N_c(N_c-1)}~
\int \frat{dq}{\pi^4} \left[T^{+T}(-q)~F^{+L~T}(q)~T^{+}(q)-
T^{+T}(q)~F^{+L}(q)~T^{+}(-q) \right]~,
$$
where $\Sigma^{+R}(p)=\Delta^{+R}~T^{+T}(-p)~\epsilon~T^{+}(p)$ and
$\Sigma^{+L}(p)=\Delta^{+L}~T^{T}(-p)~\epsilon~T(p)$.
In this paper we limit ourselves to treating the diquark condensate only.
However,
as known \cite{rapp} the mixed phase of nonzero values both chiral and colour
condensates could exist at $\mu_c\sim 300~{\mbox{MeV}}$ realizing the
transitional regime for the onset of colour superconducting phase. In order to
bring this phase to the play the equation system should be extended including
another averages as
$$
\langle\psi^{L,R}_{f\alpha i}(p)\psi^{\dagger L,R}_{g\beta j}(q)\rangle=
\delta_{fg}~\delta_{\alpha\beta}~\pi^4~\delta(p-q)~G^{LL,RR}_{ij}(p)~.
$$

From the Eqs. (\ref{2}) and (\ref{3}) we find
\begin{eqnarray}
G^{LR}(p)&=&G_0^{+}(p)+G_0^{+}(p)~\Sigma^R(p)~F^{+R~T}(p)~,
\nonumber\\
F^{+R~T}(p)&=&G^{LR~T}(-p)~\Sigma^{+L}(p)~G_0^{+}(p)~.
\nonumber
\end{eqnarray}
With the auxiliary matrices $C^{+L}(p)=\Delta^{+L}~T^{+}(-p)~T(p)$,
$C^{R}(p)=\Delta^{R}~T(p)~T^{+}(-p)$ we may rewrite the matrices $\Sigma$ as
$$\Sigma^{+L}(p)=\epsilon~C^{+L}(p)~,~~\Sigma^{R}(p)=C^{R}(p)~\epsilon~.
$$
The identities for $T$-matrix mentioned above help to show the validity of
the relations
$$\epsilon~C^{+L~T}(-p)~\epsilon^T=C^{+L}(p)~,$$
$$\epsilon~C^{R~T}(-p)~\epsilon^T=C^{R}(p)~.$$
And for the Green function we have
$$G^{LR}(p)=G_0^{+}(p)+G_0^{+}(p)~C^{R}(p)~\epsilon~G^{LR~T}(-p)~\epsilon~
C^{+L}(p)~G_0^{+}(p)~.
$$
Combining the properties of $C$ matrices together with the identities for free
Green functions
$$\sigma_2~G_0^{\pm T}(p)~\sigma_2=~G_0^{\mp}(p)~,
$$
we are able to get the complete equation for calculating the function $G^{LR}$
in the form
$$\epsilon~G^{LR}(-p)~\epsilon^T=
G_0^{-}(-p)+G_0^{-}(-p)~C^{+L}(p)~G^{LR}(p)~C^{R}(p)~G_0^{-}(-p)~.$$
The vector-function ${\vf{ h}}(p)$ being spanned by the vector ${\vf{ p}}$
helps to conclude that all the matrices $G_0^{\pm}$, $C^{+L}$, $C^R$ commute
with each other. Then searching the solution for the Green function
$G^{LR}$ by iterating one finds immediately that the Green function commutes
with those matrices and finally obtains the equation to calculate it in the
form
$$G^{LR}(p)=[1+H(p)]~G_0^{+}(p)+H^2(p)~G^{LR}(p)~,$$
or
$$[1-H(p)]~G^{LR}(p)=G_0^{+}(p)~,
$$
where $H(p)=G_0^{+}(p)~C^R(p)~G_0^{-}(-p)~C^{+L}(p)$.

The structure of matrices $H(p)$ allows us to conclude that their sum
$$H(p)+H(-p)=\alpha(p)~$$
and product
$$H(p)~H(-p)=\beta(p)~,$$
are proportional to the unity matrices (it is clear by definition that
$\alpha(-p)=\alpha(p),~\beta(-p)=\beta(p)$).

Introducing notation $g_\nu=h_\nu(-p)$ we are able to present the functions
$\alpha(p)$ and $\beta(p)$ in the compact form as
$$\alpha(p)=4~\Delta^R~\Delta^{+L}~[-4~A(p)~(hg)+2~(p^2+\mu^2)~(h^2)~(g^2)]~,
$$
where
$$A(p)=(p^2+\mu^2)~(hg)-2i~\mu~p(g_4 h-h_4 g)~.
$$
and a scalar product is naturally defined $(hg)=\sum_{i=1}^4 h_i g_i$
together with the functions $(h^2)$ and $(g^2)$ squared. Getting the last term
of $A(p)$ we used the designation of scalar function ${\vf h}$ mentioned above
and for the function $\beta$ we have
$$\beta(p)=16~\left(\Delta^R\right)^2~
\left(\Delta^{+L}\right)^2~\left(p^2+\mu^2\right)^2~(h^2)^2~(g^2)^2~.
$$

As both functions $\alpha$ and $\beta$ are spanned by the unit matrices the
solution for the Green function may be given in
$$G^{LR}(p)=\frat{G_0^{+}(p)~[1-H(-p)]}{1-\alpha(p)+\beta(p)}~.
$$
The gap equation then looks like
\begin{equation}
\label{gap}
\Delta^L=\frat{2\lambda}{N_c (N_c-1)}~\int\frat{dp}{\pi^4}~
\frat{\alpha(p)-2~\beta(p)}{\Delta^{+L}(1-\alpha(p)+\beta(p)}~.
\end{equation}
We are interested in the solution of the form
$\Delta^{R}=\Delta^{+L}=\Delta^{L}=\Delta^{+R}$ at $\lambda<0$,
which is dictated by the symmetries of four-quark interaction kernels. Let us
remember that any kernel for every quark sort traditionally carries the
imaginary unit factor $i$ \cite{2}, \cite{dp2}. The sign choice of $\lambda$
relies on this fact but, in principle, there is an alternative
$\Delta^{R}=\Delta^{+L}=-\Delta^{L}=-\Delta^{+R}$ for $\lambda>0$, if the
kernels are redefined. An analysis shows that the denominator of
Eq. (\ref{gap})
is always positive and, therefore, the solution of this equation does exist at
pretty large $\lambda$.

The quark matter state at finite chemical potential is defined by
the saddle point of functional Eq. (\ref{1}) which we treat
further maintaining the first nonvanishing contribution which is
the figure-eight type diagram (see, for example, \cite{diakcar}),
$$I=2~(N_c-1)~\int\frat{dp}{\pi^4}~
\frat{\alpha(p)-2~\beta(p)}{1-\alpha(p)+\beta(p)}~. $$ In our
consideration this contribution to the generating functional
($Z_\psi\sim e^{W}$) would occur to be
\begin{equation}
\label{GZ}
W=-n\bar\rho^4~\ln\lambda+\lambda~\langle Y\rangle~,
~~\langle Y\rangle\simeq I~.
\end{equation}
For the simplest situation of constant IL density the saddle point equation
reads
$$n\bar\rho^4-\lambda~\langle Y\rangle=0~.
$$
Comparing it with the gap equation Eq. (\ref{gap}) one could notice the
peculiar feature which is very practical to keep numerical calculations under
control. The saddle point equation leads to the condition of gap independent
of $\mu$.

It was demonstrated in Refs. \cite{we0}, \cite{we} that the quark
back-reaction upon IL could be perturbatively estimated by
studying the small variations of the IL parameters $\delta n$ and
$\delta \rho$ around their equilibrium values $n$ and $\bar\rho$.
Such variations could be incorporated by the IL theory if the
deformable (crumpled) (anti-)\-in\-stan\-tons of size $\rho$,
being the function of $x$ and $z$, i.e. $\rho\to\rho(x,z)$, are
treated as the saturating configurations of the functional
integral. Theoretically it is justified by developing the
variational principle which makes use the Ritz method based on the
multipole expansion of the deformation fields in the following
form
\begin{equation}
\label{5q} \rho(x,z)=\rho(z,z)+\frat{\partial\rho}{\partial
z_\mu}~y_\mu+\frat{\partial^2\rho}{\partial z_\mu \partial
z_\nu}~y_\mu~ y_\nu+\dots~,~~~~~y=x-z~.
\end{equation}
Then it is obvious the singular nature of the pseudoparticle
solution generates the necessary cut-off at the scale of average
instanton size.

In principle,  the instanton orientation in the colour space
should be treated in  similar manner. However, it does not lead to
noticeable consequences while in the random instanton medium.
Although such a treatment turns out very informative if one
considers the polarization characteristics of the IL when the
probe colour charge is present.

In addition, the variations of zero modes in the interaction
vertices of quark determinant at the transformation $\bar\rho \to
\bar\rho+\delta\rho$ should be taken into account. Then as the
output we have that for the long wave length excitations (for
example, $\pi$-mesons) the deformation field describes colour less
scalar excitations of IL with the mass gap $M$ of the order of
several hundreds MeV, $M^2=\frat{\nu}{\kappa}$ where
$\nu=\frat{b-4}{2}$, $b=\frat{11~N_c-2~N_f}{3}$, $\kappa$ is the
kinetic coefficient being derived within the quasi-classical
approach. $N_c$ and $N_f$ are the numbers of quark colours and
flavours, respectively. Our estimates give for this coefficient
value of a few single instanton actions $\beta=8\pi^2/g^2$, i.e.
$\kappa\sim c~\beta$ (with the factor $c\sim 1.5$ --- $6$
according to the ansatz taken for the saturating configurations)
\cite{we}.

Then, besides the diagrams with four legs (see, the term $V$ in (\ref{1}))
the extra diagrams with the scalar field attached (relatively speaking, the
derivative in $\rho$ of the vertex function in which the variation of the
functions describing the zero mode
\begin{equation}
\label{4}
h_i\to h_i+\frat{\partial h_i}{\partial \rho}~\delta\rho~,~~
i=1, \dots, 4~,
\end{equation}
should be performed) are generated. Due to the diquark condensate presence
handling the Lagrangian is substantially simplified if one
restricts oneself with the leading contributions coming from the tadpole
diagrams. Indeed, the leading contributions come from the term $V$ and the
term shown in Fig. 1 where two vertices are linked with the propagator
$\frat{1}{M^2}$ of scalar field.

\begin{figure}[htb]
\centerline{\epsfig{file=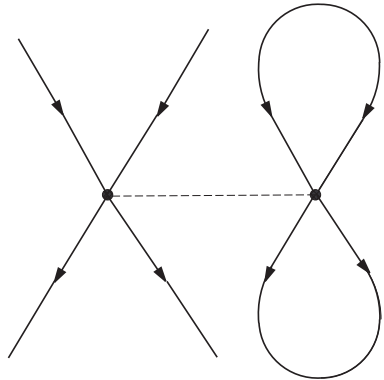,width=5.cm}} \vspace*{-.1 cm}
\caption{The diagram of tadpole approach (see, the text). The
solid (dashed) lines correspond to the fermion (scalar) field.}
\end{figure}

Analyzing the modified Lagrangian it is easy to see that the form
of the Gorkov equations would be the same if everywhere $\Sigma$
is meant as an improved one $\Sigma+\delta\Sigma$ in which the
correction $\delta\Sigma$ is constructed with the modified
functions Eq. (\ref{4}). Moreover, the result for the Green
function is formally the same if the modified functions $\Sigma$
are meant. In particular, the form of gap equation is also
retained because its kernel includes again the differences of the
matrices which are constructed from the derivatives in $\rho$ of
matrices $T,~T^{+}$ and their transposed forms. This difference is
spanned by the unit vector as well. However, the following
substitutions $$\alpha(p)\to\alpha(p)+\delta\alpha(p)~,$$
$$\beta(p)\to\beta(p)+\delta\beta(p)~,$$ where
$$\delta\alpha(p)=J~\frat{\partial \alpha}{\partial\rho}~,$$
$$\delta\beta(p)=J~\frat{\partial \beta}{\partial\rho}~,$$ are
supposed to be performed and here $J$ is the contribution of
figure-eight type diagram with the scalar field propagator as an
external leg
$$J=\frat{2\lambda}{N_c(N_c-1)}~\frat{1}{n\bar\rho^4\kappa}
\frat{1}{4 M^2}~I~. $$ Let us mention that effectively a
dependence on the kinetic term $\kappa$ disappears (remember that
$M^2=\frat{\nu}{\kappa}$) and its precise value is not operative
in the approximation developed. The derivative of the figure-eight
type diagram which will shortly be necessary to proceed looks like
$$\frat{\partial I}{\partial \rho}=2~(N_c-1)~\int\frat{dp}{\pi^4}~
\left\{\frat{1-\beta(p)}{(1-\alpha(p)+\beta(p))^2}~\delta
\alpha(p)+ \frat{\alpha(p)-2}{(1-\alpha(p)+\beta(p))^2}~\delta
\beta(p)\right\}~. $$ And finally the modification of generating
functional should be accommodated bringing Eq. (\ref{GZ}) to the
form
$$W=-n\bar\rho^4~\left(\ln\frat{n\bar\rho^4}{\lambda}-1\right)+
\lambda~\langle Y\rangle~.$$ Then the new (if the variation of the
IL density is absorbed) equation to calculate the saddle point
reads
$$n\bar\rho^4-\lambda~(n\bar\rho^4)'\ln\frat{n\bar\rho^4}{|\lambda|}-
\lambda~\langle Y\rangle=0$$ and the IL density is \cite{we0}
\begin{equation}
\label{ilden}
n\bar\rho^4=\frat{\nu}{2\beta\xi^2}+
\left[\left(\frat{\nu}{2\beta\xi^2}\right)^2+
\frat{\left(\frat{\delta I}{\delta\rho}\right)^{'}}{\beta \xi^2}~
\frat{\Gamma(\nu+1/2)}{2\sqrt{\nu}~\Gamma(\nu)}
\right]^{1/2}~,
\end{equation}
where the prime available means the derivative in $\lambda$, the constant
$\xi^2=\frat{27}{4}\frat{N_c}{N_c^{2}-1} \pi^2$
is a measure of interaction in the stochastic ensemble of PPs with an average
size as\\
$\bar\rho\Lambda=\exp\left\{-\frat{2N_c}{2\nu-1}\right\}$.
The derivative in $\lambda$ of the figure-eight type diagram looks like
$$\left(\frat{\delta I}{\delta \rho}\right)^{'}=
4~(N_c-1)\frat{\Delta^{'}}{\Delta}~
\int\frat{dp}{\pi^4}~\left\{
\frat{1+\alpha-6\beta+\alpha\beta+\beta^2}
{(1-\alpha+\beta)^3}~\delta \alpha+
\frat{-4+7\alpha+4\beta-\alpha\beta-\alpha^2}
{(1-\alpha+\beta)^3}~\delta \beta\right\}~
$$
and the derivative $\Delta^{'}$ is defined by the following equation
$$\frat{N_c (N_c-1)}{2\lambda^2}=\frat{2\Delta^{'}}{\Delta^3}
\int\frat{dp}{\pi^4}~
\frat{2\beta-2\beta^2+2\alpha\beta-\alpha^2}
{(1-\alpha+\beta)^2}~.
$$

\begin{figure}[htb]
\centerline{\epsfig{file=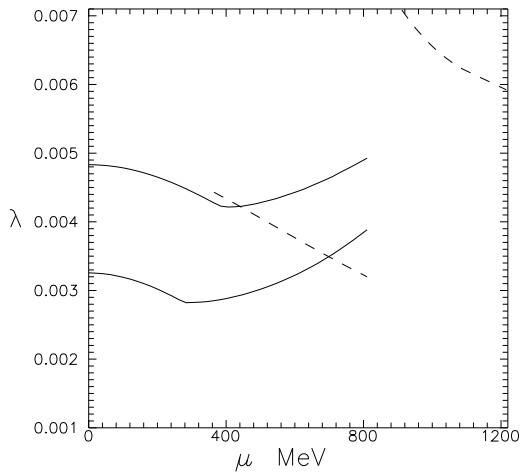,width=6.8cm}} \vspace*{-.1 cm}
\caption{The saddle point $\lambda$ as the function of chemical
potential $\mu$ at $N_c=3$ and $N_f=2$. The solid lines correspond
to the phase of broken chiral symmetry (the lower one absorbs the
IL perturbation). The dashed lines correspond to the colour
superconductivity phase (the left hand one does not include the
tadpole contribution whereas the right hand one does).}
\end{figure}

Fig. 2 shows the results of calculating the parameter $\lambda$ (which is
proportional to the free energy within the precision of one loop approximation)
as the function of chemical potential at $N_c = 3, N_f = 2.$ The dashed lines
expose a behaviour in the phase of nonzero values of diquark condensate (the
left hand line (lower line) corresponds to the calculations if the quark
interaction with
IL is ignored) whereas the solid lines show the behaviours in the phase of
broken chiral symmetry (upper line for the quark interaction with IL ignored)
 \cite{mz}. The saddle point parameter
$\lambda_1$ of Ref. \cite{mz} and that exploited in the present paper are
related as
$\lambda_1^{2}=-\frat{\lambda~n\bar\rho^4}{2(N_c-1)N_c}$.
The crossing points of the curves are fixing the onset of colour
superconducting phase. As mentioned above, actually, there is a transitional
region of mixed phase where both chiral and diquark condensates develop
nonzero magnitudes and the descent is not so steep. However, these details
are inessential for further discussion. What is more interesting to be noticed
comes from the crossing point of lower solid and dashed lines positioned
on the plot.

\begin{figure}[htb]
\centerline{\epsfig{file=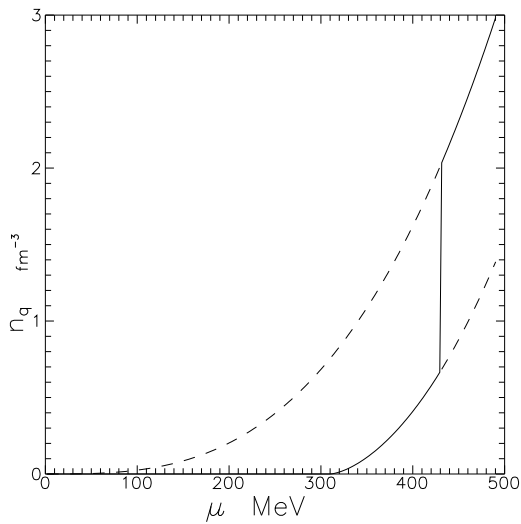,width=6.8cm}} \vspace*{-.1 cm}
\caption{The quark density being defined by the crossing point of
the upper solid and left hand dashed lines in Fig. 2. The solid
line corresponds to the stable phase and the dashed one
corresponds to the metastable one.}
\end{figure}

The first crossing point (along the $\mu$-axis) which corresponds
to the approach without the tadpole contributions included looks
to be slightly larger in magnitude than that obtained earlier
$\mu_c\simeq 300~{\mbox{MeV}}$ in Refs. \cite{diakcar},
\cite{rapp}. However, it does not signal about shifting the
critical point to larger values of $\mu$ in the approach dealing
with the initial Lagrangian. The IL parameters of various
approaches are sometimes slightly different and, in principle,
could be optimized fitting, for example, $\Lambda_{QCD}$. Thus,
the shift of $\mu_c$ value pointed out is well within a precision
of the IL theory. Regarding the second crossing point (when IL is
perturbed by quarks) the conclusion could be more indicative. Here
$\mu_c$ becomes significantly larger. The corresponding quark
matter densities as the functions of chemical potential for both
approaches are depicted in Fig. 3 and Fig. 4. Apparently, on both
plots the density values corresponding to the onset of colour
superconductivity phase are appreciably larger than the density of
normal nuclear matter. Perhaps, our result could be taken as a
general indication on the considerable role of supplementary
(insignificant on the instanton background) interactions of light
quarks. They are able to change the estimate of critical density
for colour diquark condensation drastically.

\begin{figure}[htb]
\centerline{\epsfig{file=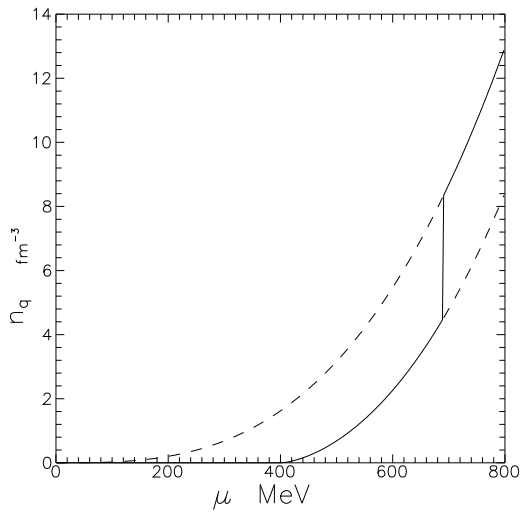,width=6.8cm}} \vspace*{-.1 cm}
\caption{The quark density being defined by the crossing point of
the lower solid and left hand dashed lines in Fig. 2. The solid
line corresponds to the stable phase and the dashed one
corresponds to the metastable one.}
\end{figure}

Now we would like to comment on the crossing point of upper dashed
and lower solid lines which is not shown in Fig. 2. It corresponds
the unrealistic values of $\mu$ and in that region the IL is
hardly applicable. One of the limits for IL approach coming from
the very large values of chemical potential points out that with
quark matter density increasing the average interquark distances
may become very small and the 'Coulomb' (perturbative) field
strengths would occur to be comparable with the (anti-)instanton
ones. Then the (anti-)instanton superposition is not a proper
configuration to saturate the functional integral. Thus, our
conclusion resulting from the approach developed gives a message
that the quark perturbation of IL leads the corresponding curves
to move apart inherent in the phases of nonzero values of diquark
and chiral condensates in Fig. 2. The 'chiral' curve becomes
steeper but 'diquark' curve is displaced to the larger values of
$\mu$ increasing its critical value. It looks like giving more
reliability to our understanding of the perturbative fields role
in the IL theory could provide us with more accurate appraisals
for the phenomena considered.

\begin{figure}[htb]
\centerline{\epsfig{file=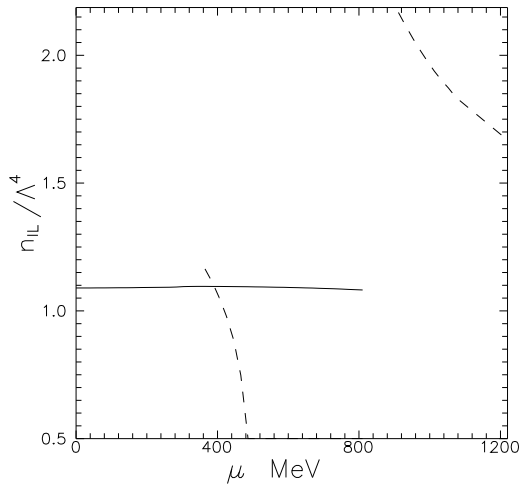,width=6.8cm}} \vspace*{-.1 cm}
\caption{ The IL density for the approach at $N_c=3$ and $N_f=2$.
The solid line corresponds to the calculation in the phase of
broken chiral symmetry. Both dashed lines correspond to the
calculation in the phase of nonzero diquark condensate values.
The lower dashed line gives the estimate of IL density without the
tadpole contributions included. Regarding the upper dashed line
see the text.}
\end{figure}

Fig. 5 demonstrates the IL density as the function of $\mu$ when
the quark back-reaction is incorporated in the phase of broken
chiral symmetry (solid curve) and in the phase of nonzero values
of diquark condensate (dashed curve). It is interesting to notice
that the curves manifest the different character of quark
influence on IL. In the phase of broken chiral symmetry the
density is almost constant in the suitable interval of $\mu$
whereas in the colour superconducting phase quickly disappears. It
is remarkable that for the former the corresponding analogue of
the tadpole contribution $\left(\frat{\delta
I}{\delta\rho}\right)^{'}$ in Eq. (\ref{ilden}) is strictly
positive and, therefore, can lead to the increase of the IL
density (gluon condensate) which is truly insignificant and
demonstrates simply the approach sensitivity to the dynamical
quark mass variation  \cite{mz}. In the latter case the sign of
tadpole contribution is changing. As seen from Fig. 5 it leads to
the drastic change of the IL density behaviour and the gluon
condensate is getting weaker. Strictly speaking this effect has
already been discussed in \cite{kerb}.

\noindent The authors are partly supported by STCU \#P015c,
CERN-INTAS 2000-349, NATO~2000-PST.CLG 977482 Grants.


\end{document}